\documentclass[aps,prd,twocolumn]{revtex4}
\usepackage{amsmath}

\begin{document}

\title{Comment on ``Quantum Tunneling Beyond Semiclassical Approximation'' by R. Banerjee and B. R. Majhi (and many others!) }
\author{Alexandre Yale}
\affiliation{Perimeter Institute, 31 Caroline St. N., Waterloo, Ontario N2L 2Y5, Canada}
\email{ayale@perimeterinstitute.ca}
\date{February 22, 2011}

\begin{abstract}
We discuss recent work which has found, using a tunneling approach, higher-order terms in the Hawking temperature.  We highlight a few important issues in the derivation, such as a misleading definition of energy, and criticize some of the conclusions that have been reached.  In particular, we conclude that contrary to many recent claims, the tunneling method yields no higher-order corrections to the Hawking Temperature.
\end{abstract}

\maketitle
\newcommand{\myeq}[1]{\begin{equation} \begin{split}  #1 \end{split} \end{equation}}

\section{Introduction}
The tunneling approach \cite{Volovik,PW,NullGeo1,PaddyTun1} is a modern technique to model Hawking radiation as the tunneling of quantum fields across an event horizon.  This method has the considerable advantage of being local, and can therefore be used to study spacetimes with multiple horizons such as embedded black holes in deSitter spacetimes.  It has therefore been used to explore the thermodynamics of many spacetimes: Kerr-Newman \cite{KN1,KN2}, Black Rings \cite{Zhao2006}, Taub-NUT \cite{Kerner2006}, AdS black holes \cite{Hemming2001}, BTZ \cite{BTZ1,BTZ2}, Vaidya \cite{Vaidya}, dynamical black holes \cite{Cri2007}, Kerr-G\"{o}del \cite{Kerner2007}, deSitter horizons \cite{dS1,dS2,dS3}, and constant curvature black holes \cite{Yale2010}.  Moreover, this approach allows quantum fields to be considered explicitly, such that the emission of scalars, fermions \cite{Fermions1,Fermions2,Fermions3,Fermions4,0809.1508,0901.2258}, and bosons \cite{Yale2010b,0901.2258} from a multitude of black holes has been studied.
\\\\
It has recently been claimed that higher-order quantum corrections can be calculated using the Hamilton-Jacobi version \cite{PaddyTun1} of the tunneling method.  This was first shown for scalars \cite{0805.2220} before being expanded to fermions \cite{0809.1508} and photons \cite{0901.2258}, finally exploding into a popular research topic \cite{guilty1,guilty2,guilty3,guilty4,guilty5,guilty6,guilty7,guilty8,guilty9,guilty10,0808.3688,1005.2264} dealing with the thermodynamics of various spacetimes.  The idea is quite simple: in the common formalism of the tunneling method, one takes the semi-classical limit $\hbar \rightarrow 0$ and retains only leading-order terms to calculate the standard Hawking temperature.  The claim of \cite{0805.2220} is that if one keeps these higher-order terms, then non-zero higher-order corrections to the Hawking temperature are found.  These corrections are later associated with back-reaction effects and a conformal trace anomaly.
\\\\
The existence of such corrections is odd for many reasons.  First, \cite{0805.2220} studies a free field on a background metric which is fixed and stationary, such that what is calculated cannot be a true back-reaction.  Moreover, the Hawking temperature can be calculated exactly through various other means, and no higher-order quantum corrections are found as long as the metric remains fixed.  Indeed, Chatterjee and Mitra \cite{Mitra} as well as Wang et al \cite{Wang2010} have recently argued that the tunneling method yields no higher-order corrections to the Hawking temperature, and it has also been shown that neither scalars, fermions, nor bosons introduce higher-order terms \cite{Yale2010b}.
\\\\
This comment aims to clarify the issue.  We will begin by summarizing the derivation from \cite{0805.2220} of higher-order terms by going over the case of the massless scalar field in Section \ref{deriv}; derivations for fermions \cite{0809.1508} and photons \cite{0901.2258} follow analogous, though progressively more complicated, steps.  In Section \ref{disc}, we highlight problems with this derivation as well as with the interpretation of its results.  In particular, we will show that this version of the tunneling method does not, contrary to previous claims, make any prediction regarding effects of back-reaction on a black hole's temperature,  and yields no higher-order corrections to the Hawking temperature.  Indeed, the calculated corrections from \cite{0805.2220} stem from a misleading definition of the energy.

\section{Summary of the derivation for a free scalar field} \label{deriv}
There exists in the literature multiple derivations of higher-order quantum corrections for the tunneling of a scalar field from the generic near-horizon black hole line element
\myeq{ ds^2 = -f(r)dt^2 + \frac{1}{g(r)}dr^2 + r^2 d \Omega^2,}
where $f(r)$ and $g(r)$ vanish at the horizon $r_0$, while $f'(r_0)$ and $g'(r_0)$ both remain finite.  Since most of these derivations are based on \cite{0805.2220}, we will reproduce here the main ideas of that paper.
\\\\
Taking a hint from the WKB approximation, the scalar field is written as $\phi = e^{\frac{-i}{\hbar} S}$ and satisfies the Klein-Gordon equation $\partial_\mu \left( g^{\mu \nu}\sqrt{-g} \partial_\nu \right) \phi = 0$.   Looking only at the $(r,t)$ sector of the equations, which contains all the divergent terms, we find
\myeq{\label{KG2}   \frac{-1}{\sqrt{fg}} \partial_t^2 \phi + \frac{1}{2} \left( f' \sqrt{ \frac{g}{f}} + g' \sqrt{\frac{f}{g}}\right)\partial_r \phi + \sqrt{fg} \partial_r^2 \phi = 0.}
The action is then expanded in powers of $\hbar$: $S = S_0 + \sum_{i=1}^\infty\hbar^i S_i$, and, setting each order to zero independently in $(\ref{KG2})$,  we find a series of equations:
\myeq{ \label{system}
&\hbar^0 \hspace{0.5cm} 0 = (\partial_t S_0)^2 - fg(\partial_r S_0)^2; \\
&\hbar^1 \hspace{0.5cm} 0 = 2i \partial_t S_0 \partial_t S_1 - 2ifg \partial_r S_0 \partial_r S_1 - \partial_t^2 S_0  \\
&\hspace{1.4cm} + fg\partial_r^2 S_0 + \frac{1}{2}(f'g+fg')\partial_r S_0; \\
&\hbar^2 \hspace{0.5cm} 0 = i(\partial_t S_1)^2 + 2i\partial_t S_0 \partial_t S_2 - ifg(\partial_r S_1)^2 \\
&\hspace{1.4cm} - 2ifg\partial_r S_0 \partial_r S_2 - \partial_t^2 S_1 + fg\partial_r^2 S_1 \\
&\hspace{1.4cm}+ \frac{1}{2}(f'g+fg') \partial_r S_1; \\
&\vdots
}
It turns out that these equations can be solved iteratively: solving for $S_i$ using $S_{j<i}$.  In the end, we find that, for every $i$,
\myeq{ \label{prop}
\partial_t S_i = \pm \sqrt{fg} \partial_r S_i
}
solves the system of equations $(\ref{system})$.  \cite{0805.2220} concludes from equation $(\ref{prop})$ that since every $S_i$ satisfies the same differential equation, they must all be proportional to one another.  In that spirit, they introduce coefficients $\beta_i$ such that
\myeq{ \label{prop2}
S_i = \frac{\hbar^i \beta_i}{M^2} S_0,
}
where the $M^2$ is introduced because we are working in units where $\hbar$ has dimensions of mass squared, since we set $c = G = k_B = 1$.  Following the steps of the standard zeroth-order calculation, they define the energy by $E = -\partial_t S_0$ and find
\myeq{ \label{action}
\text{Im} S &= \text{Im} \oint \partial_r S = \text{Im} \oint \frac{ E \left( 1 +  \sum_i \frac{\hbar^i \beta_i}{M^{2i}} \right)}{\sqrt{fg}} \\
&= \left( 1 +  \sum_i \frac{\hbar^i \beta_i}{M^{2i}} \right) \text{Im} S_0  ;
}
this allows them to find corrections to the Hawking temperature through the tunneling rate \footnote{Since \cite{0805.2220} has been published, there have been considerable advances in our understanding of the tunneling method, as summarized in \cite{Gill2010} and \cite{Akhmedov2008}.  Here, we use the simplified equation $(\ref{tempdef})$ because it agrees with the proper modern method in this case, while saving us the trouble of having to introduce a number of concepts ultimately irrelevant to the problem at hand.} $\Gamma \propto e^{-2 \text{Im} S} = e^{-E/T}$:
\myeq{\label{temp}
T = T_0 \left( 1 +  \sum_i \frac{\hbar^i \beta_i}{M^{2i}} \right) ^{-1}.
}
Similar derivations were later performed for fermions \cite{0809.1508} and photons \cite{0901.2258}.

\section{Discussion} \label{disc}
In disagreement with the above, it has been shown that neither scalars \cite{Mitra,Wang2010} nor fermions or bosons \cite{Yale2010b} lead to higher-order corrections.  We will discuss this contradiction by outlining problems with the above derivation and highlighting where the results have been misinterpreted by the authors.
\\\\
First, because this method does not fix the coefficients $\beta_i$, equation $(\ref{temp})$ is vacuous: it merely states that the temperature consists of a bare part, $T_0$, plus some unspecified correction.  In particular, it can be trivially derived without doing any actual calculations, as long as we assume that the emission rate goes as $\Gamma \propto e^{-E / T}$ where $E=-\partial_t S_0$ \footnote{We will see later in this section that this definition for the energy is misleading; we use it here because it is the definition used to derive equation $(\ref{temp})$, which we are trying to rederive without having to even introduce field equations.} as in the previous section.  Indeed, to leading order, the tunneling rate goes as $\Gamma_0 \propto e^{-2 \text{Im} S_0} = e^{-E_0/T_0}$, whereas to the next order, it goes as $\Gamma_1 \propto e^{-2 \text{Im} (S_0 + \hbar S_1) } = e^{-E_0/T}$.  Combining these two relations leads to the equality $T \text{Im} \left(  S_0 + \hbar S_1 \right)= T_0 \text{Im} S_0$, which in turn implies $T = T_0 \left( 1 + \hbar \frac{\text{Im} S_1}{\text{Im} S_0} \right)^{-1}$.  Since $\text{Im} S_0$ and $\text{Im} S_1$ have already been evaluated to correspond to the path of the emitted field (as in, for example, equation $(\ref{action})$), they are merely numbers.  For example, for the case of a Schwarzschild black hole, $2 \text{Im} S_0 = 8 \pi M E$.  Denoting the ratio between these two numbers by $\frac{\beta_1 \hbar}{M^2}$, we retrieve equation $(\ref{temp})$.
\\\\
Second, and most importantly, these results do not imply higher-order corrections to the Hawking temperature, and the higher-order terms from equation $(\ref{temp})$ are the result of a misleading definition of the energy.  Indeed, equation $(\ref{prop2})$ says that all the $S_i$ are proportional to one another; in particular this means that
\myeq{ \label{action2}
S = S_0 \left( 1 +  \sum_i \frac{\hbar^i \beta_i}{M^{2i}} \right).
}
Because the tunneling rate is given by $\Gamma \propto e^{-2 \text{Im} S} = e^{-E/T}$, the temperature can easily be computed:
\myeq{ \label{tempdef}
T = \frac{E}{2 \text{Im} S} = \frac{E}{2 \text{Im} S_0 \left( 1 +  \sum_i \frac{\hbar^i \beta_i}{M^{2i}} \right)}.
}
It is clear from this expression that the temperature is directly dependent on our definition of the energy.  In a stationary spacetime like the one we are considering, a sensible definition for the energy is $E = -\partial_t S$; using equations $(\ref{action2})$ and $(\ref{tempdef})$, this directly implies
\myeq{
T = \frac{-\partial_t S_0}{2 \text{Im} S_0} = T_0.
}
The left hand side is the temperature $T$ as calculated to every order in $\hbar$, while the right-hand side is the zeroth-order Hawking temperature $T_0$.  Therefore, there are no higher-order quantum corrections to the Hawking temperature.  These claimed corrections, as seen in equation $(\ref{temp})$, are simply the result of defining the energy by $E = -\partial_t S_0$ instead of $E = -\partial_t S$.  Because it is possible \cite{Yale2010b} to perform all the calculations exactly without ever expanding in powers of $\hbar$, it makes no sense to define the energy from the zeroth-order part of the action (which may not even be well-defined) as $E = -\partial_t S_0$, just as it would make no sense to define it as $E = \partial_t (S/2)$.  It is therefore extremely misleading to claim that the higher-order terms in equation $(\ref{temp})$ are due to higher-order effects, as this hides the fact that the concepts of temperature and energy go hand in hand.  When considering higher-order terms in the action, the definition of energy must be updated analogously, and the non-trivial result is that the Hawking temperature sees no correction.
\\\\
We should note, however, that back-reaction does have interesting effects on Hawking radiation.  Indeed, Hawking \cite{Hawking} initially calculated one-loop corrections to the radiation process, which were associated with a trace anomaly.  Later, the tunneling method was used to show that this back-reaction modified the thermal nature of the emitted radiation \cite{PW}.  More recently, it was used to study correlations between emitted particles, which may provide an answer to the information loss paradox \cite{Zhang2009,Israel2010}.

\section{Conclusion}
Contrary to the claims of \cite{0805.2220} and many subsequent papers \cite{guilty1,guilty2,guilty3,guilty4,guilty5,guilty6,guilty7,guilty8,guilty9,guilty10,1005.2264,0809.1508, 0901.2258, 0808.3688}, the tunneling method yields no higher-order corrections to the Hawking temperature.   In particular, for the Schwarzschild black hole, this means that, \emph{to every order in $\hbar$}, the Hawking temperature is
\myeq{ T = \frac{\hbar}{8 \pi M} .}
This is because we are working with a fixed metric and a free field, and therefore obviously disregard back-reaction effects.  The higher-order quantum corrections found in  \cite{0805.2220} are merely a mathematical artifact of a misleading definition of energy, and not a consequence of higher-order physical effects.  Our arguments are in line with those of Chatterjee and Mitra \cite{Mitra} as well as Wang et al \cite{Wang2010}, and it is our hope that this will put an end to the deluge of papers claiming to calculate higher-order corrections to various spacetimes using this technique.

\acknowledgments{
The author would like to thank Doug Singleton for insightful comments regarding the results presented here.  This work was supported by the Natural Sciences and Engineering Research Council of Canada.
}

\end{document}